\documentstyle[12pt,epsfig]{article}             %%%
\newcommand{\beq}{\begin{equation}}             %%
\newcommand{\eeq}{\end{equation}}               %%
\newcommand{\bqry}{\begin{eqnarray}}            %%
\newcommand{\eqry}{\end{eqnarray}}              %%
\newcommand{\bqryn}{\begin{eqnarray*}}          %%
\newcommand{\eqryn}{\end{eqnarray*}}            %%
\newcommand{\preprint}[1]{\begin{table}[t]      %%
            \begin{flushright}                  %%
            \begin{large}{#1}\end{large}        %%
            \end{flushright}                    %%
            \end{table}}                        %%
\newcommand{\PD}[2]                             %%
    {\frac{\partial^{#2}}{\partial #1^{#2}}}    %%
\begin{document} 
\preprint{LA-UR-02-2690} 
\title{An Analytic Model of the Gr\"{u}neisen Parameter \\ at All Densities} 
\author{\\ Leonid Burakovsky\thanks{E-mail: burakov@lanl.gov} \
and Dean L. Preston\thanks{E-mail: dean@lanl.gov}
 \\  \\ 
%Theoretical Division, MS B283 \\  
Los Alamos National Laboratory \\ Los Alamos, NM 87545, USA }
\date{ }
\maketitle
\begin{abstract}
\vspace*{0.1cm}
\hspace*{-0.75cm}
We model the density dependence of the Gr\"{u}neisen parameter as 
$\gamma (\rho )=1/2+\gamma _1/\rho ^{\;\!1/3}+\gamma _2/\rho ^{\;\!q},$ where 
$\gamma _1,$ $\gamma _2,$ and $q>1$ are constants. This form is based on 
the assumption that $\gamma $ is an analytic function of $V^{1/3},$ and was 
designed to accurately represent the experimentally determined low-pressure 
behavior of $\gamma $. The numerical values of the constants are obtained 
for 20 elemental solids. Using the Lindemann criterion with our model for 
$\gamma ,$ we calculate the melting curves for Al, Ar, Ni, Pd, and Pt and 
compare them to available experimental melt data. We also determine the $Z$ 
(atomic number) dependence of $\gamma _1.$ The high-compression limit 
of the model is shown to follow from a generalization of the Slater, 
Dugdale-MacDonald, and Vashchenko-Zubarev forms for the dependence 
of the Gr\"{u}neisen parameter.
\end{abstract}
\bigskip

\section*{Introduction} 

Lattice anharmonicity leads to a volume dependence of the phonon frequencies, 
$\omega_i$, that is described by the mode Gr\"{u}neisen parameters \cite{Gru}
\beq
\gamma _i=-\frac{\partial \ln \omega _i}{\partial \ln V},
\eeq
and the mechanical Gr\"{u}neisen parameter is defined as the average of the 
$\gamma _i$ over the first Brillouin zone, $\gamma _m\equiv \langle \gamma _i
\rangle .$ When the $\gamma _i$ are all equal  it can be shown \cite{Gir} that 
$\gamma _m = \gamma _i$ coincides with the thermodynamic Gr\"{u}neisen 
parameter \cite{Bar} 
\beq
\gamma _{th}=\frac{\alpha VB_T}{C_V},
\eeq
where $\alpha $ is the thermal expansion coefficient, $B_T$ is the isothermal 
bulk modulus, and $C_V$ is the heat capacity at constant volume. If the 
$\gamma _i$ are not all equal, $\gamma _m\neq \gamma _{th}$ in general. As 
noted by Barron \cite{Bar}, Born has shown that, even if the $\gamma _i$ are 
not all equal, $\gamma _m=\gamma _{th}$ in the limit of low $(T\rightarrow 0)$ 
and high $(T\stackrel{>}{\sim }\Theta _D)$ temperatures. The $T=0$ limiting 
value of $\gamma _m=\gamma _{th}\equiv \gamma $ is $\gamma =-d\ln \Theta _D/d
\ln V,$ where $\Theta _D$ is the Debye characteristic temperature, $\hbar 
\langle \omega_i^{-3}\rangle ^{-1/3}/k_B $, in the limit $T\rightarrow 0$ 
\cite{Bar}. Here both $\Theta _D$ and $\gamma $ are functions of volume, or 
density, alone. This formula is the Debye-Gr\"{u}neisen definition of $\gamma $ 
\cite{Poi}, which can be rewritten in terms of density, $\rho \sim 1/V,$ as 
\beq 
\gamma (\rho ) = \frac{d\ln \Theta_D(\rho )}{d\ln \rho}.
\eeq

The Lindemann melting criterion, which asserts that the 
root-mean-square atomic displacement of atoms from their equilibrium 
positions in a solid is a fixed fraction of the interatomic distance 
at the melting point, can be rewritten in the form of the Gilvarry law, 
which relates the density derivative of the melting temperature, 
$T_m(\rho )$, to the Gr\"{u}neisen parameter \cite{Gil}: 
\beq 
\frac{d\ln T_m(\rho )}{d\ln \rho}=2\left[ \gamma (\rho )-\frac{1}{3}\right] ,
\eeq
hence integration of $\gamma (\rho )$ yields the melt curve. Equations of 
state for solids can be constructed on the basis of $\gamma (\rho )$, though 
additional thermodynamic data are needed \cite{JD}. 

There is a long history of attempts to model $\gamma (\rho)$ \cite{ZK}.  
Slater \cite{SL}, Dugdale and MacDonald \cite{DM}, and Vashchenko and 
Zubarev \cite{VZ} proposed three expressions for $\gamma (\rho )$ that 
are summarized by the single formula \cite{formula} 
\beq 
\gamma =\frac{\frac{B'}{2}-\frac{1}{6}-\frac{t}{3}\left( 1-\frac{P}{3B}
\right) }{1-\frac{2t}{3}\;\!\frac{P}{B}}, 
\eeq
where $P$ is pressure on the cold curve ($T = 0$ equation of state),
and $B\equiv -V\;\!dP/dV$ and $B'=(dB/dV)/(dP/dV)$ are 
the bulk modulus and its pressure derivative at $T = 0$. We will refer to 
Eq.\ (5) as the SDMVZ (Slater, Dugdale-MacDonald, Vashchenko-Zubarev) formula.
Slater's derivation is based on the Debye-Gr\"{u}neisen definition (3) and 
assumes no volume dependence of the Poisson ratio. Dugdale and MacDonald 
used a simplification of lattice dynamics in which the material is modeled 
as a lattice undergoing one-dimensional harmonic oscillations. Their 
derivation was improved by Vashchenko and Zubarev who considered 
three-dimensional oscillations of a lattice with interatomic interactions 
described by an anharmonic central potential. Equation (5) reduces to the 
Slater, Dugdale-MacDonald, and Vaschenko-Zubarev formulas for $t=0,$ 1, 
and 2, respectively. Eq.\ (5) has since been rederived by other researchers 
following different approaches (\cite{IS,BS,WL}).
 
\section*{An analytic model for $\gamma (\rho )$}

Melting curves and equations of state are usually based on simple functional 
forms for $\gamma (\rho )$. Dongquan and Wanxing \cite{DW}, for example, 
proposed the form $\gamma (\rho )= \gamma _0\rho _0/\rho +2/3\;(1-\rho _0/\rho
 )^\delta $, which reduces to $\gamma \rho \approx {\rm constant}$ at small 
compressions and approaches 2/3 as very large compressions; $\gamma $ does not 
differ from that given by Eq.\ (5) by more than 10\% for $1\leq \delta \leq 3$ 
\cite{DW}. For $\delta =2$ this form reduces to $\gamma (\rho )=2/3\;\!+\;\!
\gamma _1/\rho \;\!+\;\!\gamma _2/\rho ^2$, a model frequently used for the 
construction of equations of state and melting curves \cite{JD,YC}.
 
At very high compressions, of order 10, a solid becomes a crystallized 
one-component plasma, i.e., a lattice of ions in a uniform neutralizing 
background of electrons \cite{Young}. Several theoretical studies predict 
$\gamma =1/2$ for this limiting state of a solid.  Kopyshev \cite{K} 
calculated $\gamma (V)$ in the Thomas-Fermi approximation and found $\gamma 
\rightarrow 1/2$ as $V\rightarrow 0.$ Simple dimensional arguments by Hubbard 
\cite{Hu} also indicate that $\gamma \rightarrow 1/2$ as $V\rightarrow 0$. 
Additional theoretical studies that give $\gamma \rightarrow 1/2$ include, 
but are not limited to, references \cite{R,NN,MWYZ,Pe}. Nevertheless, some 
researchers assume $\gamma \rightarrow 2/3$ as $V\rightarrow 0$ \cite{HHS} 
because that is the limiting value of $\gamma $ as given by Eq.\ (5) with 
fixed $t$ (see below). 

We now construct a simple, practical three-parameter model of $\gamma (\rho)$ 
that accurately fits low-compression data, is equivalent to the experimental 
result that $\gamma \rho ^{\;\!q_{\rm eff}}\approx {\rm constant}$, $q_{\rm 
eff}\geq 1$, for compressions up to 1.5, and limits to $1/2$ as $V\rightarrow 
0$. We assume that the Gr\"{u}neisen parameter is an analytic function of $x
\equiv V^{1/3}$, essentially the interatomic distance, and that the coefficient 
of $x$ in the Taylor-Maclaurin series expansion for $\gamma $ is non-zero. 
The simplest model that satisfies all of these requirements is 
\beq
\gamma (V)=\frac{1}{2}+c_1V^{1/3}+c_2\;\!V^q,\;\;\;c_1,c_2,q={\rm const ,}
\;\;q>1,
\eeq
Note that $\gamma (V)-1/2$ is asymptotic to $c_1V^{1/3}$; this is discussed 
in more detail in the next section. The term $c_2\;\!V^q$ represents the 
contribution of the quadratic and higher-order terms in $x$ which must sum 
to make $\gamma $ a concave-up function of $V$ at small compressions. 

Eq.\ (6) can be rewritten in terms of density as
\beq
\gamma (\rho )=\frac{1}{2}+\frac{\gamma _1}{\rho ^{\;\!1/3}}+\frac{\gamma _
2}{\rho ^{\;\!q}},\;\;\;\gamma_1,\gamma _2,q={\rm const ,}\;\;q>1,
\eeq
Note that both $\gamma _1$ and $\gamma _2$ are dimensional parameters. 

Eq.\ (7) incorporates the low-compression power-law behavior 
$\gamma \rho ^{\;\!q_{\rm eff}}\approx {\rm constant}$, $q_{\rm eff}\geq 1$, 
known from experiment. Experimental values of $q_{{\rm eff}}$ are typically 
in the range 1 to 2 \cite{A}, but may be as high as 3 \cite{AIY}. The effective 
exponent can be defined by $q_{{\rm eff}}=-d\ln \gamma /d\ln \rho $ since 
$\gamma \sim \rho ^{-q_{\rm eff}}$; in our model we have $q_{{\rm eff}}=
[\gamma _1/(3\rho ^{1/3})+q\gamma _2/\rho ^q]/\gamma (\rho )$. Averaging 
over compressions from 1 to 1.25 we find $\langle q_{{\rm eff}}\rangle 
\leq 3$ in general. Consider, for example, gold: with the corresponding 
parameters from Table 1, $q_{{\rm eff}}$ varies from 4.7 at $\rho =19.3$ 
g/cc to 1.3 at $\rho =24$ g/cc with an average value of 2.7.

With $\gamma (\rho )$ given by Eq.\ (7), the Lindemann equation (4) can be 
integrated to yield the melting curve
\beq
T_m(\rho )=T_m(\rho _r)\left( \frac{\rho }{\rho _r}\right) ^{
1/3}\exp \left\{ 6\gamma _1\left( \frac{1}{(\rho _r)^{1/3}}
-\frac{1}{\rho ^{\;\!1/3}}\right) +\frac{2\gamma _2}{q}\left( 
\frac{1}{(\rho _r)^q}-\frac{1}{\rho ^{\;\!q}}\right) \right\} ,
\eeq
where $\rho _r$ and $T_m(\rho _r)$ are a reference density and corresponding 
melting temperature. 

We now demonstrate that the asymptotic form of our model for the Gr\"{u}neisen 
parameter follows from a generalization of the SDMVZ formula. 

\section*{Asymptotic $\gamma (\rho )$ from a generalization of the SDMVZ 
formula}

Parshukov \cite{Pa} studied the compression dependence of $\gamma $ for lead, 
indium, and tin in the pressure range 0.6 to 4.0 GPa and found that it is 
almost identical for all three metals but is not described by Eq.\ (5) over the 
entire range of pressures for $t$ a constant. He found that the Slater formula 
$(t=0)$ provided the most accurate fit at small compressions while the 
Dugdale-MacDonald one $(t=1)$ is optimal at higher compressions. He suggested 
that all three $(t=0,1,2)$ formulas be unified by making $t$ compression 
dependent. Romain {\it et al.} \cite{RMJ2} noted that the melting curve of 
Al cannot be described in the Lindemann approach using the $t=0,$ 1, or 2 
formulas for $\gamma (\rho ).$ Nagayama and Mori \cite{NM} analyzed available 
experimental data on 16 metals and found that the data on $\gamma $ at $P=0$ 
are best described by the Slater formula, but at moderate compressions the 
data are best fit by the Dugdale-MacDonald formula which is approximately 
equivalent to $\gamma /V^{1.1}={\rm constant,}$ close to the relation $\gamma 
/V={\rm constant}$ that is routinely used in high-pressure studies. Nagayama 
and Mori noted that $d\ln \gamma /d\ln V$ decreases with increasing 
compression from the value given by the Slater formula ($t=0$) through that 
given by the Dugdale-MacDonald formula ($t=1$) toward the value predicted 
by the Vashchenko-Zubarev relation ($t=2$). An analysis of Hugoniot data shows 
that $t$ follows a similar trend $(0\rightarrow 1\rightarrow 2)$ in Al, Cu, and 
Ta \cite{Wu}.  These analyses of experimental data indicate that the parameter 
$t$ is not a constant but a variable that increases with increasing compression.

Irvine and Stacey \cite{IS} generalized the Vashchenko-Zubarev formula by 
accounting for non-central interatomic forces. They added a term $f$ to the 
expression for the interatomic force constant and obtained the formula
\beq
\gamma = \frac{\frac{B'}{2}-\frac{5}{6}+\frac{2P}{9B}-\frac{f}{18B}+
\frac{1}{6}\;\!\frac{df}{dP}}{1-\frac{4P}{3B}+\frac{f}{3B}}, 
\eeq
which reduces to the Vashchenko-Zubarev formula for $f = 0$. By introducing 
the dimensionless parameter $t$ via 
\beq
f=2P\;\!(2-t),
\eeq
Eq.\ (9) reduces to 
\beq
\gamma = \frac{\frac{B'}{2}-\frac{1}{6}-\frac{t}{3}\left( 1-\frac{P}{3B}
\right) +\frac{P}{3B}\;\!V\;\!\frac{dt}{dV}}{1-\frac{2t}{3}\;\!\frac{P}{B}}, 
\eeq
which would coincide with Eq.\ (5) without the additional term $(PV/3B)\;dt/dV$ 
in the numerator. Hence Eq.\ (11) is an extension of the SDMVZ formula, Eq.\ 
(5), to the case of density-dependent $t$. We note that Eq.\ (11) can be cast 
in the form 
\beq
\gamma =\frac{1}{2}\;\!\frac{B}{B-\frac{2}{3}tP}\;\!\frac{d\left( B-\frac{
2}{3}tP\right) }{dP}-\frac{1}{6}=-\;\!\frac{1}{2}\;\!\frac{d\ln \left( B-
\frac{2}{3}tP\right) }{d\ln \;\!V}-\frac{1}{6},
\eeq
which is equivalent to Eq.\ (5) if $t$ is a constant. In view of the 
experimental evidence that $t$ is a decreasing function of $V$ we use 
Eq.\ (11) rather than (5) for our subsequent analysis.

We consider Eq.\ (11) at ultrahigh pressures where the ($T = 0$) 
equation of state is accurately given by \cite{B,Ho}
%\beq
%P=aV^{-5/3}-cV^{-4/3},\;\;\;a,c={\rm const}>0,
%\eeq
\beq
P=aV^{-5/3}e^{-bV^{1/3}},\;\;\;a,b={\rm const}>0.
\eeq
Eq.\ (13) includes an exponential screening correction to the 
equation of state of a free electron gas, namely $P=aV^{-5/3}$ where 
$a=2.337\;Z^{\;\!5/3}$ TPa $\stackrel{\;\circ }{A}$$^5,$ $Z$ being 
the atomic number \cite{Ho}. It agrees with very accurate numerical 
Thomas-Fermi-Dirac results to within 2\% over the compression range of 
1 to 15 with the exception of the alkali and alkaline-earth metals \cite{B}. 
Using the equation of state (13) in Eq.\ (11) we obtain
\beq
\gamma =\frac{4\;\!(5-2\;t)+2\;\!(4-t)\;b\;V^{1/3}+b^{2}\;V^{2/3} + 6\;V\;
dt/dV}{6\;\!(5-2\;t+b\;V^{1/3})}.
%\gamma =\frac{1}{3}\;\!\frac{2a\;\!(2t-5)-3c\;\!(t-2)V^{1/3}}{
%a\;\!(2t-5)-2c\;\!(t-2)V^{1/3}}.
\eeq
It follows that $\gamma \rightarrow 2/3$ as $V\rightarrow 0$ for every 
asymptotic value of $t$ {\it except} $t=5/2$.  If $t=5/2$ then $\gamma 
\rightarrow 1/2$ as $V\rightarrow 0,$ in agreement with theoretical 
predictions. We conclude that $t$ is always asymptotic to $5/2$. Recent 
computer calculations of the compression dependence of $t$ for $\gamma $-Fe 
\cite{BS} corroborate this conclusion and also show that $t$ can saturate at 
quite moderate compressions: $t=2.46\pm 0.03$ at compressions of only 1.5 
to 2.
  
The cold-curve ($T=0$) pressure (13) is an analytic function of $x$ everywhere 
except at the origin (infinite compression) where it has a pole of order five. 
(The majority of model equations of state are analytic in $x$ except for poles 
at $x=0$. See Holzapfel \cite{Ho} for a list of 16 examples.) Consequently, 
both $P/B$ and $B'$ are analytic for $x\geq 0$. Given the analyticity of 
$\gamma ,$ $P/B,$ and $B'$ for $x\geq 0$, it follows from Eq.\ (11) that $t$ 
must also be an analytic function of $x$ for $x\geq 0,$ hence it can be 
represented by the power series 
\beq
t=5/2-\sum_{i=1}^{\infty }t_i\,x^i.
\eeq
The first sub-leading coefficient, $t_i$, must be non-negative if $t$ is in 
fact a monotonically increasing function of compression. Substituting this 
series in Eq.\ (14) and expanding we find that the asymptotic form of $\gamma $ 
is  
\beq
\gamma =\frac{1}{2}+\frac{b^2+2\,b\,t_1-2\,t_2}{6\,b+12\,t_1}\;\!V^{1/3}+
\ldots \;\;.
\eeq
It is expected that the $V^{1/3}$ term is always present in the series 
expansion of $\gamma $, i.e., $b^2+2\,b\,t_1-2\,t_2 \neq 0$. Thus the 
generalized SDMVZ formula leads to $\gamma \sim 1/2+c_1x,\;\!x\rightarrow 0$, 
in agreement with Eq.\ (6). 
%As the contribution of the higher-order 
%terms can be effectively represented by the term $c_2V^q,$ $q>1,$ one finally 
%arrives at Eq.\ (6) for $\gamma (V).$ 

In summary, we have (i) generalized the SDMVZ formula for $\gamma (\rho )$ 
to account for a density-dependent $t$, (ii) shown that $\gamma $ goes to its 
theoretical high-pressure limit $1/2$ only if $t$ is asymptotic to $5/2$, and 
(iii) demonstrated that our model for the Gr\"{u}neisen parameter and the 
generalized SDMVZ formula, Eq.\ (11), have the same asymptotic behavior 
provided $t$ is an analytic function of $x$, and $t(0)=5/2$. 

\section*{Determination of model parameter values}

In this section we outline the procedure for extracting the numerical 
values of the three model parameters $\gamma_1$, $\gamma_2$, and $q$ 
from data and determine their values for 20 elemental solids.

Ideally, one would fit our model (7) to data on $\gamma (\rho )$ and extract 
the values of $\gamma _1,\gamma _2$, and $q$, but such data sets are very 
rarely available. Alternatively, one could fit the functional form (8) to 
$T_m(\rho )$ data, but such data are very rarely available either since 
experiment usually determines $T_m(P)$. Typically, the only available data 
on $\gamma $ are $\gamma (\rho _{300} =\rho (P=0,\;\!T=300$ $^\circ $K)), 
which is approximately equal to $\gamma _{th}(T = 300$ $^\circ $K), and 
$\gamma (\rho_m=\rho (P=0,\;\!T=T_m))$ which can be found by equating 
$d T_m / d \rho$ at $P = 0$ as given by the Kraut-Kennedy law 
\cite{K-K}
\beq
T_m(\rho )=T_m(\rho _m)\left[ 1+2\left( \gamma (\rho _m)-\frac{1}{3}\right) 
\left( \frac{\rho }{\rho _m}-1\right) \right] 
\eeq
to $dT_m/d\rho =(B_m/\rho _m)\;\!dT_m/dP$, where $B_m$ and $dT_m/dP$ at 
$P=0$ are taken from experiment. The value of $B_m$ is obtained by calculating 
bulk moduli at low temperatures from measured single-crystal elastic constants 
and then extrapolating to $T_m$. The pressure derivative $dT_m/dP$ is obtained 
either directly from low-pressure melting curve data or indirectly from 
isobaric-heating measurements of $\triangle H$ ($H$ is the enthalpy) and 
$\triangle V$ across the melting transition so that $dT_m/dP=T_m\triangle V/
\triangle H$ (Clausius-Clapeyron equation). Eq.\ (17) is obtained by expanding 
$T_m(\rho )$ in a power series about $\rho _m$ using Eq.\ (4). Equation (7) 
with $\rho =\rho_{300}$ and $\rho =\rho_m$ provides two conditions needed to 
determine the three parameters $\gamma _1,$ $\gamma _2$, and $q$. The third 
condition comes from the ultrahigh pressure limit, which we discuss next.

The melting curve of a solid at ultrahigh pressures is described by the 
equation
\beq
\frac{Z^2e^2}{ak_BT_m}=\Gamma _m,
\eeq
where $a=(3v/4\pi )^{1/3}$ is the Wigner-Seitz radius $(v$ being the 
Wigner-Seitz volume) and $\Gamma _m$, a dimensionless constant, is the OCP 
coupling parameter at melt. The value of the coupling parameter is $170-180$ 
for a body-centered cubic (bcc) OCP crystal \cite{BP} (the recent calculation 
of $\Gamma_m$ for a bcc crystal by Potekhin and Chabrier \cite{PC} gave 
$\Gamma_m=175.0\pm 0.4)$, and as high as $200-210$ for a face-centered cubic 
(fcc) OCP crystal \cite{HCV,Ree}. In the following analysis we make no 
distinction between bcc and fcc OCP crystals and take $\Gamma _m=180.$ 

It follows from Eq.\ (18), and Eq.\ (8) in the limit $\rho \rightarrow 
\infty $ with $\rho_r = \rho_m$ 
that 
\beq
T_m(\rho _m)\;\!v_m^{1/3}\exp \left\{ \frac{6\gamma _1}{\rho _m^{1/3}}+\frac{
2\gamma _2}{q\rho _m^q}\right\} =\left( \frac{4\pi }{3}\right) ^{1/3}\frac{
e^2}{k_B}\;\!\frac{Z^2}{\Gamma _m},
\eeq
where $v_m$, the Wigner-Seitz volume at the zero-pressure melting point, 
equals $a_m^3/2$ for a bcc and $a_m^3/4$ for a fcc crystal; $a_m$ is the 
lattice constant. 
%In Eq.\ (22), the numerical value of $e^2/k_B$ is $1.671\cdot 10^{-3}$ 
%$^\circ {\rm K}\cdot {\rm cm}.$ 

We obtain the values of the parameters $\gamma _1,$ $\gamma _2$ and $q$ by 
simultaneous solution of three non-linear equations, namely Eq.\ (7) for 
$\rho =\rho_{300}$ and $\rho =\rho_m$, and Eq.\ (19) with $\Gamma_m = 180$. 
The value of $\gamma (\rho _{300}),$ which is very rarely measured 
directly, can be approximated by $\gamma_{th}(300)=\alpha B_T/(\rho C_V)$ 
where $\alpha ,$ $B_T,$ and $C_V$ are all measured at 300~$^\circ $K; 
the necessary room-temperature data can be found in ref.\ \cite{web}, 
and the data on $\gamma _{th}(300)$ in ref.\ \cite{GK}. Parameter values 
for 20 elements are shown in Table 1; $\gamma _1$ and $\gamma _2$ are 
given to three significant figures, and $q$ to two significant figures.

The uncertainty in the value of $\gamma (\rho _m)$ is determined 
by the error bars on the measured melting temperatures and densities, usually 
a few percent (except perhaps for laser-heated diamond anvil cell (DAC) 
melt data, see below). Typical uncertainties in $\alpha $, $B_T,$ and $C_V$ 
result in an uncertainty of roughly 5\% in the value of $\gamma (\rho _{300})$ 
(we consider the difference between $\gamma (\rho _{300})$ and $\gamma_{th}(
300)$ to be negligible). These uncertainties in $\gamma (\rho _m)$ and $\gamma 
(\rho _{300})$ imply an uncertainty $\sim 5$\% in $\gamma _1$, values of $q$ 
that are accurate to $\sim 40$\%, and order-of-magnitude uncertainties in 
$\gamma _2$ that are attributable to a very strong sensitivity to the value 
of $q$. Let us take, as an example, copper; with $\gamma (\rho_m) = 2.48$ 
and $\gamma_{th}(300)=2.19\pm 0.1$ we find, neglecting uncertainties in the 
density, $\gamma _1=1.84\pm 0.08,$ $\gamma _2=430-1.95\cdot 10^6,$ and 
$q=4.7\pm 2.0.$ 

Although the values of $\gamma _2$ have order-of-magnitude uncertainties, we 
note that if Eq.\ (7) is rewritten in a ``scaled'' form, $\gamma =1/2+\tilde{
\gamma }_1(\rho_{300} /\rho)^{1/3}+\tilde{\gamma }_2(\rho_{300} /\rho)^q,$ 
the values of both $\tilde{\gamma }_1$ and $\tilde{\gamma }_2$ are of order 1, 
as can be seen in Table 1 where we also include the values of both $\tilde{
\gamma }_1$ and $\tilde{ \gamma }_2$ to three significant figures.
\\

%\vspace*{0.2cm}
%\begin{table}[t]
\begin{center}
{\footnotesize
\begin{tabular}{|l|l|c|l|c|c|c|c|c|l|}
\hline %-----------------------------------------------------------------------
{\rm element} & $Z$ & $\gamma _1,$ $\left( \frac{{\rm g}}{{\rm cc}}\right) $$^{
1/3}$ & $\gamma _2,$ $\left( \frac{{\rm g}}{{\rm cc}}\right)$$^q$ & $\tilde{
\gamma }_1$ & $\tilde{\gamma }_2$ & $q$ & $\rho _{300},$ g/cc & $\rho _m,$ 
g/cc & $T_m(\rho _m),$ $^\circ $K   \\
\hline %-----------------------------------------------------------------------
 Ne (fcc) & 10 & 1.04 & 5.17                & 0.91 & 2.10 & 2.2 & 1.507 & 
 1.435 & 24.6    \\
 Mg (hcp) & 12 & 0.79 & 4.33                & 0.66 & 0.53 & 3.8 & 1.740 & 
 1.640 & 923     \\
 Al (fcc) & 13 & 0.84 & 45.4                & 0.60 & 1.40 & 3.5 & 2.700 & 
 2.550 & 933.5   \\
 Ar (fcc) & 18 & 1.19 & 16.3                & 0.90 & 1.86 & 3.8 & 1.771 & 
 1.622 & 83.8    \\
% V (bcc)  & 23 & 1.39 & 74.9               & 3.0 &       & 5.630 & 2183    \\
 Fe (bcc) & 26 & 1.72 & $2.66\cdot 10^3$    & 0.86 & 0.25 & 4.5 & 7.870 & 
 7.270 & 1811    \\
 Co (fcc) & 27 & 1.81 & $6.28\cdot 10^4$    & 0.88 & 0.39 & 5.5 & 8.830 & 
 8.180 & 1768    \\
 Ni (fcc) & 28 & 1.85 & $5.60\cdot 10^5$    & 0.89 & 0.38 & 6.5 & 8.900 & 
 8.220 & 1728    \\
 Cu (fcc) & 29 & 1.87 & $2.31\cdot 10^4$    & 0.90 & 0.78 & 4.7 & 8.930 & 
 8.370 & 1357.7  \\
 Zn (hcp) & 30 & 1.91 & $1.84\cdot 10^3$    & 0.99 & 1.05 & 3.8 & 7.140 & 
 6.900 & 692.7   \\
 Mo (bcc) & 42 & 2.06 & $1.40\cdot 10^6$    & 0.95 & 0.19 & 6.8 & 10.21 & 
 9.650 & 2896    \\
 Pd (fcc) & 46 & 2.40 & $3.34\cdot 10^6$    & 1.05 & 0.25 & 6.6 & 12.02 & 
 11.29 & 1828    \\
 Ag (fcc) & 47 & 2.23 & $9.63\cdot 10^4$    & 1.02 & 1.21 & 4.8 & 10.49 & 
 9.850 & 1235.1  \\
 Cd (hcp) & 48 & 2.43 & $2.47\cdot 10^4$    & 1.18 & 0.97 & 4.7 & 8.650 & 
 8.420 & 594.2   \\
 In (bct) & 49 & 2.43 & $6.14\cdot 10^3$    & 1.25 & 0.80 & 4.5 & 7.310 & 
 7.200 & 429.8   \\
 Sn (bct) & 50 & 2.37 & $2.37\cdot 10^3$    & 1.22 & 0.83 & 4.0 & 7.300 & 
 7.200 & 505.1   \\
% Ta(bcc) & 73 & 2.06 & $5.16\cdot 10^7$    & 7.0 &       & 15.40 & 3290    \\
% Ta(bcc) & 73 & 2.82 & $1.05\cdot 10^6$    & 6.7 &       & 15.40 & 3290    \\
% W (bcc)  & 74 & 2.90 & $1.30\cdot 10^8$   & 7.0 & 19.25 & 17.63 & 3695    \\
 Pt (fcc) & 78 & 3.21 & $1.13\cdot 10^{11}$ & 1.16 & 1.01 & 8.3 & 21.45 & 
 20.19 & 2041    \\
 Au (fcc) & 79 & 3.21 & $1.97\cdot 10^{12}$ & 1.20 & 1.62 & 9.4 & 19.30 & 
 18.29 & 1337.6  \\
 Tl (bcc) & 81 & 3.17 & $3.74\cdot 10^7$    & 1.39 & 0.54 & 7.3 & 11.85 & 
 11.55 & 577     \\
 Pb (fcc) & 82 & 3.09 & $8.21\cdot 10^8$    & 1.38 & 0.89 & 8.5 & 11.34 & 
 11.05 & 600.6   \\
 U (bcc)  & 92 & 3.39 & $1.05\cdot 10^{11}$ & 1.27 & 0.57 & 8.8 & 19.05 & 
 17.60 & 1405    \\
\hline %-----------------------------------------------------------------------
\end{tabular}
}
%\caption{ }
\end{center}
%\end{table}
%\vspace*{0.25cm}
Table 1. Numerical values of the parameters entering Eqs.\ (7) and (8) for 
20 elemental solids. The crystal structure indicated for an element is that 
from which it melts at zero pressure. For Ne and Ar the entry under $\rho _{
300}$ is the value of $\rho (T=0)$.
% \\

\section*{Melting curves}

In this section we calculate melting curves using Eq.\ (8) with parameters 
determined in the previous section and compare the results with the available 
experimental data. 

In Fig.\ 1 we compare our theoretical aluminum melting curve to experimental 
data \cite{HL}. Although the theoretical melting curve looks like a fit to 
the datapoints, the parameters $\gamma _1,$ $\gamma _2$, and $q$ were {\it not} 
obtained from such a fit but rather from the zero-pressure data $\gamma (\rho 
_{300}=2.7$ g/cc)=2.5 \cite{GK}, $\gamma (\rho _m=2.55$ g/cc)=2.83 \cite{K-K}, 
and $T_m(\rho _m)=933.5$~$^\circ $K, and Eqs.\ (7) and (19). In Figs.\ 2-4 we 
compare our theoretical melting curves for argon, nickel, palladium, and 
platinum to melting data available in the literature. Agreement between 
our theoretical curves and experiment is good with the exception of the 
lower-pressure data on nickel which were obtained from a laser-heated DAC 
\cite{LS}. This discrepancy may be due to systematic errors in the DAC data 
\cite{Dai}. 

\section*{The $Z$ dependence of $\gamma _1$}

It is evident from Table 1 that $\gamma _1$ is a slowly increasing function 
of atomic number. We have fit the forms $\gamma _1=C_1\cdot Z^{\;\!n}+C_2$ 
and $\gamma _1=C_1\cdot (\ln Z)^{\;\!n}+C_2,$ where $C_1,$ $C_2$ and $n$ are 
constants, to the table entries and find that the minimum $\chi ^2,$ 0.11, 
is realized for both $\gamma _1=(12/11)\;Z^{1/3}-11/7$ and $\gamma _1=(2/21)\;(
\ln Z)^{7/3}+2/13$. These fits may be used to predict $\gamma _1$ in those 
cases where data provide only two constraints on $\gamma _1$, $\gamma _2$ and 
$q$. In Fig.\ 5 we plot $\gamma _1=(2/21)\;(\ln Z)^{7/3}+2/13$ along with the 
$\gamma _1(Z)$ entries in Table 1. 

We note that the $\tilde{\gamma }_1$ entries in Table 1 can be fitted to the 
same functional form, i.e., $\tilde{\gamma }_1(Z)=(7/26)\;Z^{1/3}+7/59$, but 
the accuracy of the fit is much lower than that of the $\gamma _1(Z)$ fit: 
$\chi ^2=0.97$.

\section*{Concluding remarks}

Our model for $\gamma $ accurately fits low-pressure data and agrees with 
theoretical predictions that $\gamma \rightarrow 1/2$ at ultrahigh pressures. 
Its accuracy cannot be determined at intermediate compressions because there 
are no experimental data. However, comparison of our calculated melting 
curves for aluminum and argon to the corresponding data in Figs.\ 1 and 2 
demonstrates that our model is accurate up to compressions of at least 2. 
(Our calculated melting curve for copper, not discussed in this paper, is 
in excellent agreement with a new SESAME copper melting curve \cite{JD} up to 
compressions $\sim 10.)$ In our model the deviation of $\gamma \rho ^{\;\!q_{
\rm eff}}$ from a constant over the range of compressions from 1 to 1.5 is 
generally less than 10\%. Finally, we note that our model for $\gamma $ 
can be used to calculate shear moduli along the solidus using Eq.\ (8) and 
the formula $G(\rho ,T_m(\rho ))/(\rho \cdot T_m(\rho ))={\rm constant }$ 
\cite{BPS}. 

\section*{Acknowledgements} 

We wish to thank J.C. Boettger, C.W. Greeff, J.D. Johnson, and G.W. Pfeufer 
for helpful discussions on the subject of the Gr\"{u}neisen parameter. One of 
us (L.B.) wishes to thank V.N.~Zharkov for very useful correspondence on the 
analytic form for the Gr\"{u}neisen parameter, and S.V.~Stankus for valuable 
data on the densities of solids at their normal melting points.

\newpage
%\hskip 1.9in    
% \epsfysize=3in
\begin{center}
%\vspace{-0.3cm} 
%\parbox{6in}{abs } 
\epsfig{file=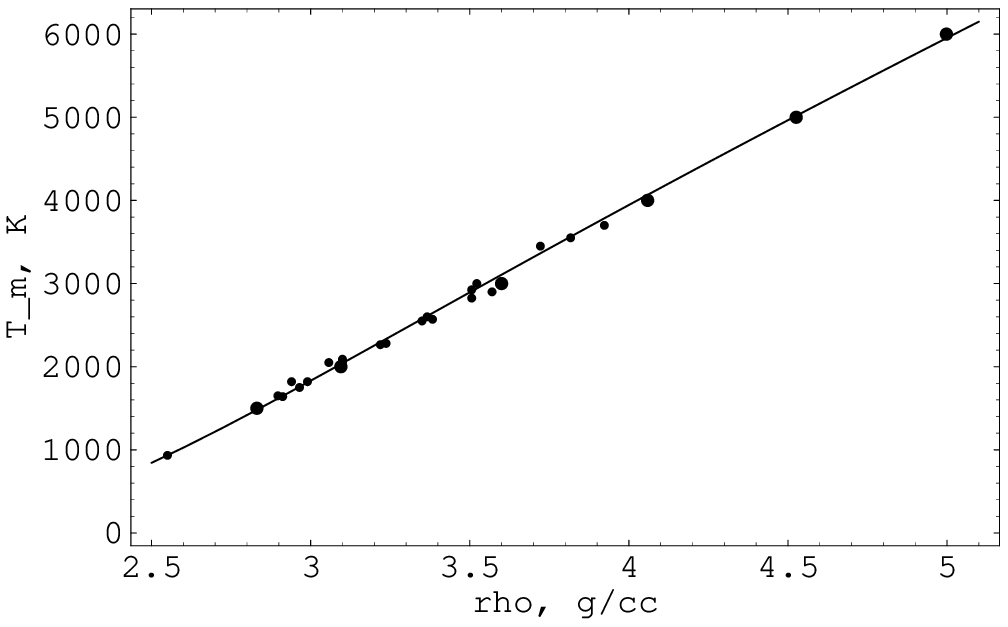,width=15cm,angle=0}
\end{center} 
\centerline{Fig.\ 1}
%. Melting curve of Al: Eq.\ (8) with the Al parameters from Table 1 
%vs. the experimental data \cite{HL} and the theoretical data \cite{MYR} 
%(thicker points). The experimental points have error bars of $\sim 100-150$ 
%$^\circ $K which are not shown on the plot.
% \\
\vspace*{1.5cm}

%\newpage
%\hskip 1.9in    
% \epsfysize=3in
\begin{center}
%\vspace{-0.3cm} 
%\parbox{6in}{abs } 
\epsfig{file=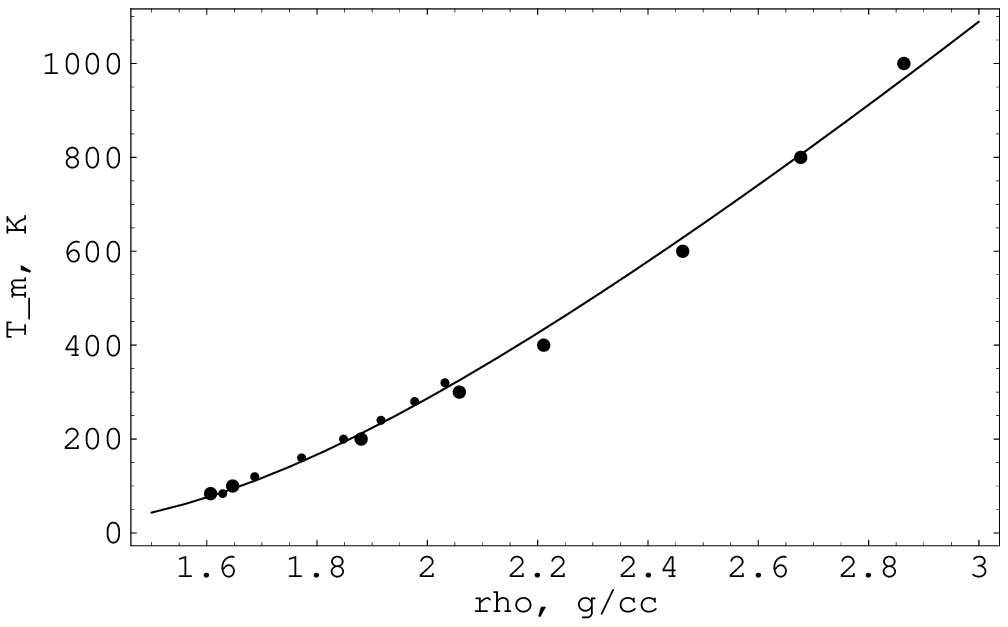,width=15cm,angle=0}
\end{center} 
\centerline{Fig.\ 2}
%. Melting curve of Ar: Eq.\ (8) with the Ar parameters from Table 1 
%vs. the experimental data \cite{CDC} and the theoretical data \cite{ZBYR} 
%(thicker points).
% \\
\vspace*{1.0cm}

%\newpage
%\hskip 1.9in    
% \epsfysize=3in
\begin{center}
%\vspace{-0.3cm} 
%\parbox{6in}{abs } 
\epsfig{file=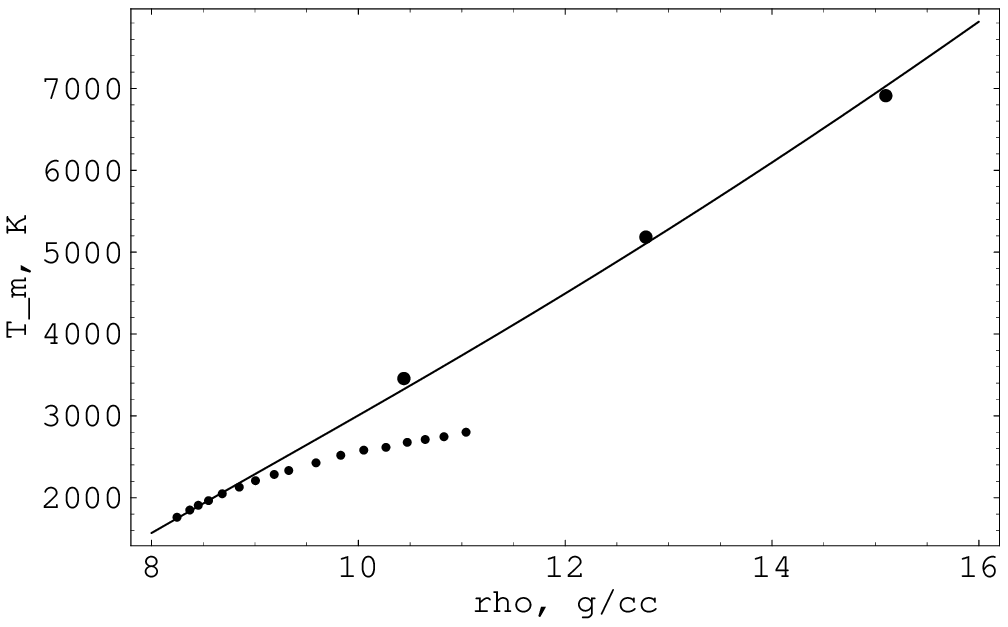,width=15cm,angle=0}
\end{center} 
\centerline{Fig.\ 3}
%. Melting curve of Ni: Eq.\ (8) with the Ni parameters from Table 1 vs. 
%the experimental data \cite{LS}. The upper three points are shock-melting 
%data from ref.\ \cite{U} which does not quote error bars. 
% \\
\vspace*{1.5cm}

%\newpage
%\hskip 1.9in    
% \epsfysize=3in
\begin{center}
%\vspace{-0.3cm} 
%\parbox{6in}{abs } 
\epsfig{file=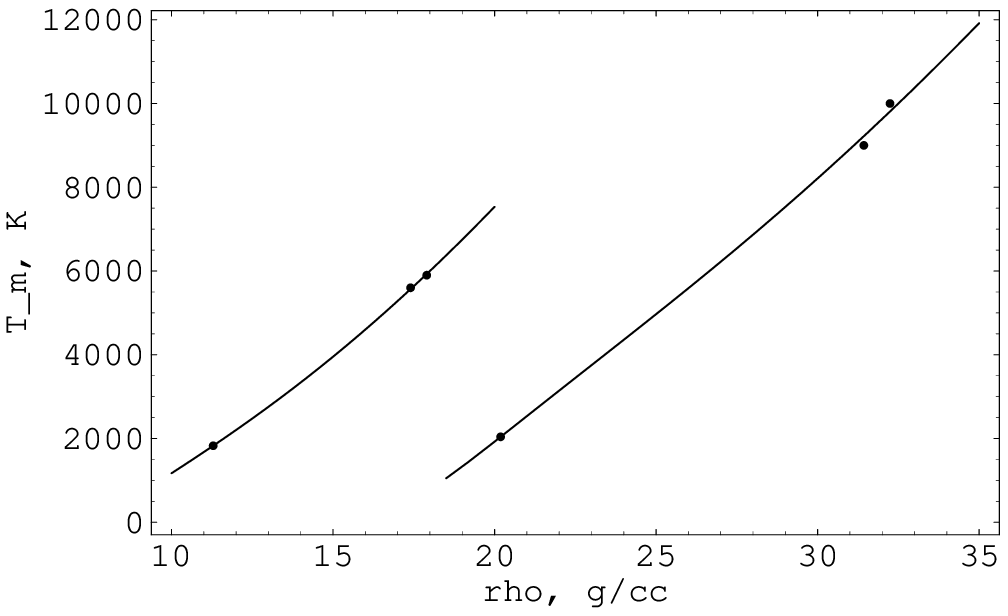,width=15cm,angle=0}
\end{center} 
\centerline{Fig.\ 4}
%. Melting curves of Pd and Pt: Eq.\ (8) with the Pd and Pt parameters 
%from Table 1 vs. shock-melting data \cite{JC} (upper two points of each 
%curve, no error bars given) and zero-pressure datapoints (lowest points on 
%each curve).
% \\
\vspace*{1.0cm}

%\newpage
%\hskip 1.9in    
% \epsfysize=3in
\begin{center}
%\vspace{-0.3cm} 
%\parbox{6in}{abs } 
\epsfig{file=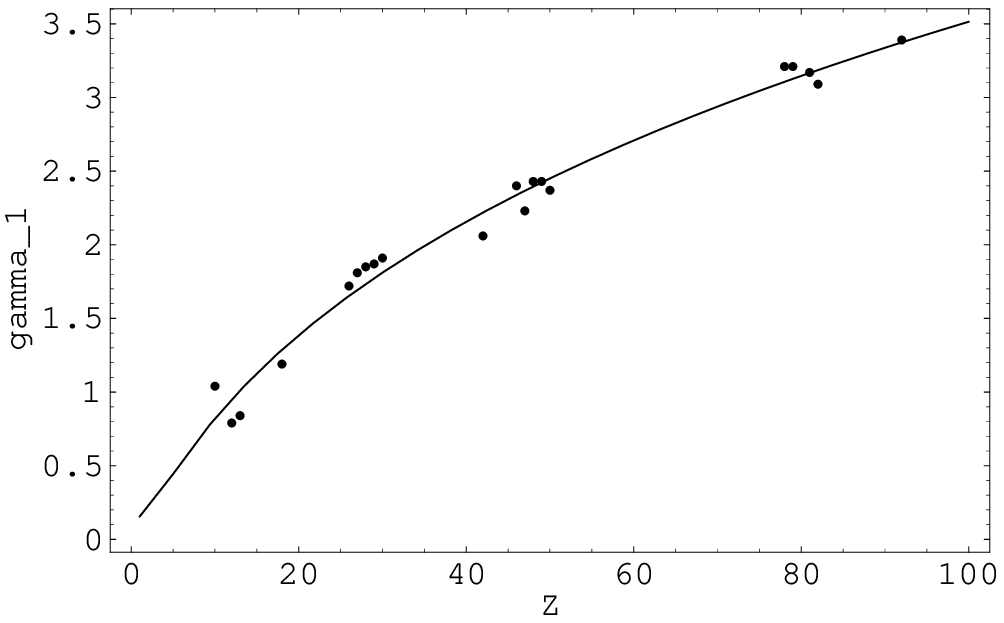,width=15cm,angle=0}
\end{center} 
\centerline{Fig.\ 5}
%. Comparison of $\gamma _1=(2/21)\;(\ln Z)^{7/3}+2/13$ to the $\gamma_1(Z)$ 
%entries in Table 1.
% \\
\vspace*{1.0cm}

%\newpage
\section*{Figure captions}

Fig.\ 1. Melting curve of Al: Eq.\ (8) with the Al parameters from 
Table 1 vs.\ the experimental data \cite{HL} and the results of 
calculations \cite{MYR} (larger points). The experimental points have 
error bars of $\sim 100-150$ $^\circ $K which are not shown on the plot.
 \\  \\
Fig.\ 2. Melting curve of Ar: Eq.\ (8) with the Ar parameters from 
Table 1 vs.\ the experimental data \cite{CDC} and the results of 
calculations \cite{ZBYR} (larger points).
 \\  \\
Fig.\ 3. Melting curve of Ni: Eq.\ (8) with the Ni parameters from 
Table 1 vs.\ the experimental data \cite{LS}. The upper three points are 
shock-melting data from ref.\ \cite{U} which does not quote error bars. 
 \\  \\
Fig.\ 4. Melting curves of Pd and Pt: Eq.\ (8) with the Pd and Pt 
parameters from Table 1 vs.\ shock-melting data \cite{JC} (upper two 
points of each curve, no error bars given) and zero-pressure datapoints 
(lowest points on each curve).
 \\  \\
Fig.\ 5. Comparison of $\gamma _1=(2/21)\;(\ln Z)^{7/3}+2/13$ to the 
$\gamma_1(Z)$ entries in Table 1.
% \\
 
\newpage 
%\bigskip
%\bigskip


\begin{thebibliography}{9}
\bibitem{formula} This formula seems to have first appeared in ref.\ [10], 
in the form $$\gamma =\frac{t-2}{3}\;\!-\;\!\frac{V}{2}\;\!\frac{d^2(PV^{2t/
3})/dV^2}{d(PV^{2t/3})/dV},$$ which is equivalent to (5) for constant $t.$
\end{thebibliography}
\end{document}